\documentclass[prl,twocolumn,showpacs,floatfix]{revtex4-1}
\pdfoutput=1
\usepackage[usenames,dvipsnames]{color}
\usepackage{longtable}
\usepackage{epsfig}
\usepackage{amssymb}
\usepackage{amsmath}
\usepackage{verbatim}
\usepackage{xspace}
\usepackage{titlesec}
\titleformat{\section}{\normalfont\fontsize{10}{15}\bfseries\filcenter}{\thesection}{1em}{}

\include{jdefs}
\definecolor{darkblue}{rgb}{0,0,0.5}
\definecolor{darkgreen}{rgb}{0.1,0,0.3}
\definecolor{darkred}{rgb}{0.6,0,0}
\usepackage{graphics}
\usepackage{hyperref}

\newcommand{\ba}{\begin{eqnarray}}
\newcommand{\ea}{\end{eqnarray}}
\newcommand{\be}{\begin{equation}}
\newcommand{\ee}{\end{equation}}

\newcommand{\weq}{{C_{\mu\nu\sigma\tau}}C^{\mu\nu\sigma\tau}}
\newcommand{\rmunu}{ R_{\mu\nu} }

\newcommand{\NAG}{\texttt{NAG} }
\newcommand{\mqd}{m^{2}}

\newcommand{\sdm}{S_{2-}}
\newcommand{\sdp}{S_{2+}}

\begin{document}
\title{Characterizing black hole metrics in quadratic gravity}
\author{A.~Bonanno${}^{a,b}$ and S. Silveravalle${}^{c}$}

\affiliation{
\mbox{${}^a$ INAF, Osservatorio Astrofisico di Catania, via S.Sofia 78, I-9    5123 Catania, Italy  }\\
\mbox{${}^b$ INFN, Sezione di Catania,  via S. Sofia 64, I-95123,Catania, Italy.}\\
\mbox{${}^c$ Dipartimento di Fisica G. Occhialini, Universit\`a degli studi di Milano Bicocca} \\
\mbox{Piazza della Scienza, 3, 20126 Milano, Italy}
}


\begin{abstract}
The recent discovery of non-Schwarzschild  black hole spacetimes has opened new directions of research in higher-derivative gravitational theories. However, despite intense analytical and numerical efforts, the link with the linearized theory
is still poorly understood. In this work we address this point for the Einstein-Weyl Lagrangian, whose weak field limit is 
characterized by the standard massless graviton and a spin-2 ghost. We show that
the strength of the Yukawa term at infinity
determines the thermal properties of the black hole and the structure of the singularity near $r=0$.
Moreover, inspired by recent results in the Asymptotic Safety scenario we investigate the consequences of an imaginary ghost mass. 
In this case we find a countable set of solutions all  
characterized  by spatial oscillations of typical wavelength determined by the mass of the spin-2 field. 
\end{abstract}

\maketitle
\section{Introduction}
Since the seminal work by Stelle \cite{stelle77}  it has long been known that extensions of Einstein's gravitational theory containing $R^2$ and  $\weq$ operators in the Lagrangian are renormalizable in $d=4$ dimensions, but at the price of a loss of unitarity owing to a spin-2 ghost. 
In recent times the possibility of defining the continuum limit around a non-Gaussian fixed point \citep{nieder}  has fueled new interest in this problem.  Several authors have in fact  proposed possible solutions for  
the unitarity problem \citep{anselmi17,piva,benedetti09}, although the physical content of the theory is still not completely understood.

At the classical level, the Einstein-Weyl theory has recently attracted considerable interest. 
It is defined by the following action
\be\label{action}
S=\int d^4 x \sqrt{-g} (\gamma R -\alpha \weq)
\ee
where  $C_{\mu\nu\sigma\tau}$ is the Weyl tensor. According to a generalized Israel theorem \cite{lu15}, any static black-hole solution of a generic quadratic gravity theory  must have vanishing Ricci scalar $R$ in the exterior region (provided $\partial_r R$  goes to zero sufficiently rapidly at spatial infinity) and then be a solution of (\ref{action}).
This property has greatly simplified the search for black hole solution in quadratic gravity.
In fact solutions  with vanishing Ricci scalar and non-vanishing Ricci tensor $R_{\mu\nu}$  have been first discovered by \cite{lu15} and further investigated in \cite{lu15b,gold18,podo}. 

However, due to its strongly non-linear nature, the general problem of connecting the weak-field  regime with the strong   one, is still not completely clarified. 
This is a central issue to discuss possible phenomenological and astrophysical implications of quadratic gravity. For instance if in the linearized theory the spin-2 mode is tachyonic as it emerges in the context of the Asymptotic Safety scenario \cite{machado,hamada}, 
the Yukawa-like behaviour $(1/r) e^{\pm m r}$ of the fields at large distances is turned into a periodic hair of the type $\sim (1/r) \cos( | m | r)$ and the spacetime is no longer asymptotically flat. One would then like to know how the properties of the horizon (\textit{i.e.} its location, surface gravity) and the further interior evolution, are determined by the asymptotic fields.

In this work, in order to tackle this problem, we employ  a multiple shooting approach which allows a complete characterization of the solutions from $r=\infty$ down to $r=0$.  Depending of the black hole ``mass'' $M$ defined at large distances we shall see   that the thermal properties of the black holes and the behavior of the metric coefficient near $r=0$ are  determined by the Yukawa coupling at infinity.  
On the contrary, for $m^2<0$  we discover a countable number of families of solutions, representing a new type of black hole. In particular the metric coefficient show characteristic ripples 
of wavelength  $\sim 1/|m|$ which resembles a gravitational 
analogous of the Friedel oscillations in plasma \cite{friedel, boos18}.


\begin{figure}[htpb]
\includegraphics[width=.48\textwidth]{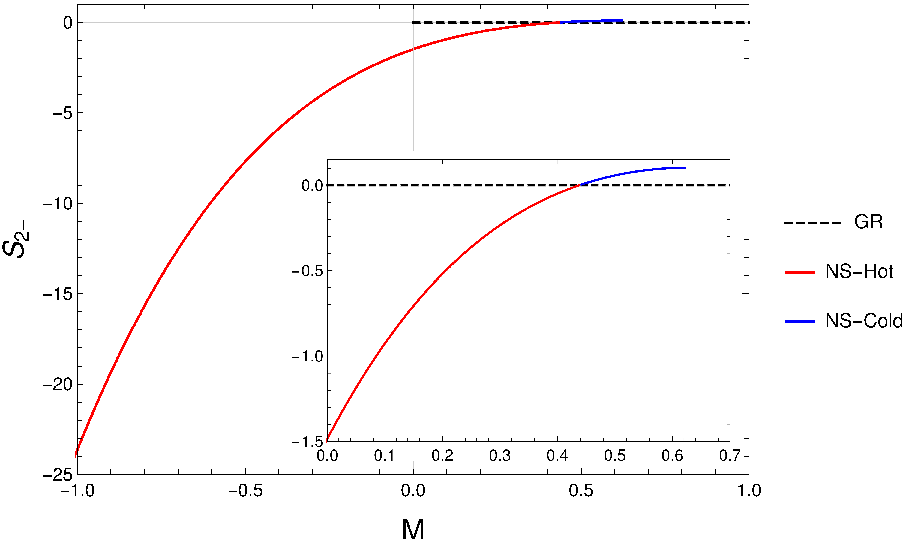}
\caption{Solutions space of Non-Schwarzschild BHs. Red line indicates non-Schwarzshild black holes
which are hotter than  Schwarzshild black holes, blue line indicates colder black holes which 
are instead obtained for $S_{2-}>0$.}
\label{fig1}
\end{figure}

\section{Linearized solutions and numerics}\label{sec1}
Let us consider the  field equations for (\ref{action})
\be\label{eom}
H_{\mu\nu}= \rmunu-\frac{1}{2} R g_{\mu\nu}-\frac{2}{\mqd}(\nabla^\rho\nabla^\sigma+\frac{1}{2}R^{\rho\sigma})C_{\mu\rho\nu\sigma}
\ee
where $m^2=\gamma/2\alpha$, and assume  a static spherically symmetric spacetime of the form
\be\label{metric}
ds^2 = -h(r) dt^2 +\frac{dr^2}{f(r)} + r^2 d\Omega^2 .
\ee
Let us then write  
\begin{equation}
h(r)=1+V(r),\quad f(r)=1+W(r)
\end{equation}
with  $V\ll 1$ and $W\ll 1$. 
The linearized field equations, describing the metric at large distances, can easily be obtained from the equations ${H_\mu}^\mu=0$ and ${H_0}^0-{H_i}^i=0$, which reduce to
\begin{subequations}
\begin{equation}
\nabla^2\left(\nabla^2V(r)+2Y(r)\right)=0
\end{equation}
\begin{equation}
\left(\nabla^2-\frac{3}{2}m^2\right)\nabla^2V(r)-\nabla^2Y(r)=0
\end{equation}
\end{subequations}
where $Y(r)=r^{-2}\left(rW(r)\right)'$ (see \cite{stelle77} for a general discussion on the weak field expansion in quadratic gravity).
The form of the linearized solution is 
\begin{equation}
\label{ag}
\begin{split}
&h(r)=1+C_t-\frac{2M}{r}+2 \sdm\frac{\mathrm{e}^{-m r}}{r}+2 \sdp\frac{\mathrm{e}^{m r}}{r}\\
&f(r)=1-\frac{2M}{r}+\sdm\frac{\mathrm{e}^{-m r}}{r}(1+m r)\, +\\
&+\sdp\frac{\mathrm{e}^{m r}}{r}(1-m r)
\end{split}
\end{equation}
where the dependence on the unknown $C_t$, $M$, $\sdp$ and $\sdm$ is explicit. 
Standard time parametrization at $r=\infty$ implies $C_t=0$  and
asymptotically flat solutions have $\sdp=0$. 
The key question we are interested in is the determination of the values of $M$ and $\sdm$ 
for which a BH solution with a non vanishing Ricci scalar is obtained.\\  
\begin{figure}[htpb]
\includegraphics[width=.48\textwidth]{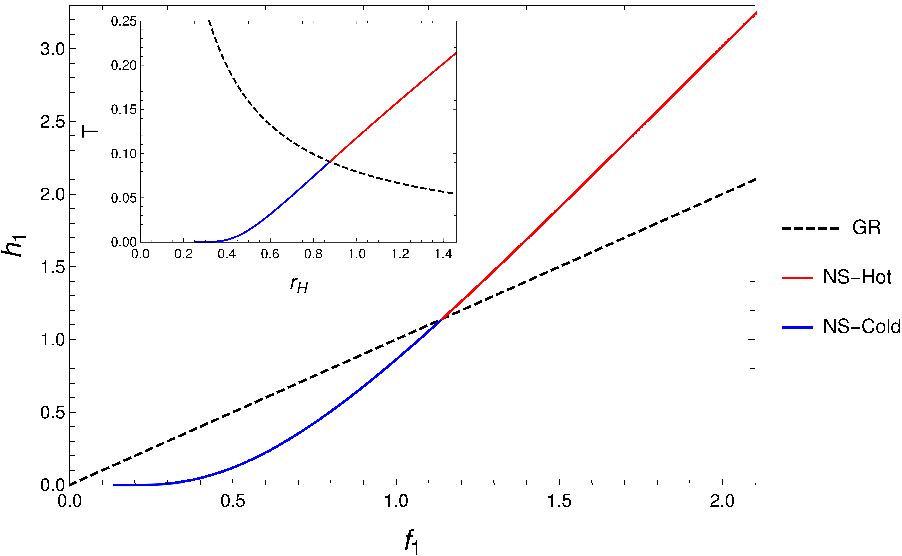}
\caption{Near-horizon parameters of the black hole solutions described in Fig.(\ref{fig1}) and Hawking temperature.}
\label{fig2}
\end{figure}
The trace of field equations (\ref{eom}) implies $R=0$ which,
upon using the spherically symmetric ansatz eq.(\ref{metric}), 
leads to the following second order equations:
\ba
\label{uno}
&&0=r h(r) \left(r f'(r) h'(r)+2 f(r) \left(r h''(r)+2 h'(r)\right)\right)+\nonumber \\
&&4 h(r)^2 \left(r f'(r)+f(r)-1\right)-r^2 f(r) h'(r)^2 
\ea
An additional second order equation can be obtained by considering a suitable combination \cite{perkins} 
\be
H_{rr}-X(r) {H_\mu}^\mu-Y(r)({H_\mu}^\mu)^2-Z(r)\partial_r {H_\mu}^\mu=0
\ee
After some manipulations one obtains
\ba
\label{due}
&&0=-r^2 f(r) h(r) \left(r f'(r)+3 f(r)\right) h'(r)^2\\
&&+2 r^2 f(r)h(r)^2 h'(r) \left(-r f''(r)-f'(r)+2 \mqd r\right)\nonumber\\
&&+h(r)^3 (r \left(-3 r f'(r)^2+4 f'(r)-4 \mqd r\right)  +r^3 f(r)^2 h'(r)^3\nonumber\\
&&+4 f(r) \left(r^2 f''(r)-r f'(r)+\mqd r^2+2\right)-8 f(r)^2)\nonumber
\ea
It is not difficult to show that near a horizon $r_H$
the following local expansion holds
\begin{subequations}\label{localh}
\ba
f(r)= f_1 (r-r_H)+f_2(f_1,r_H)\, (r-r_H)^2+...\qquad\
\ea
\ba
h(r)= h_1\left((r-r_H)+h_2(f_1,r_H)\, (r-r_H)^2+...\right)\quad\,
\ea
\end{subequations}
where $f_1$ and $h_1$ are two free parameters and $f_{i>1}$, $h_{i>1}$ are completely determined by $f_1$ and $r_H$, so that the Schwarzschild solution is obtained for $f_1=h_1=1/r_H$.
In particular the surface gravity is given by 
\ba
\kappa=\frac{1}{2}\sqrt{f_1 h_1}
\ea  
and, as always, the Hawking temperature is $T=\kappa /2\pi$.

The space of the possible black hole solutions can be obtained by means of the 
following numerical strategy.   

A weak field solution is  assumed to be valid starting from some radius $r\gg 1/m$, where the 
initial conditions (\ref{ag}) are set, and  an inward numerical integration of (\ref{uno}-\ref{due})  towards a fitting radius $r_f>r_H$
is performed. In particular, the rising Yukawa exponentials are switched off
and  ($M$, $S_{2-}$) are assumed to be arbitrary.
Moreover,   $h_1$ and $f_1$  in (\ref{localh})
determine the initial condition for a corresponding outward integration towards $r_f$ from $r_H$.
For actual calculations the Adaptive Stepsize Runge-Kutta integrator DO2PDF 
implemented by the \NAG group (see \url{https://www.nag.com} for details) turned out to be rather
efficient. 
Continuity at the fitting point of the functions $f(r)$, $h(r)$ 
and their derivatives 
as a function of  ($M$, $S_{2-}$, $h_1$, $f_1$) is obtained by means
of a globally convergent Broyden's method as described in 
\url{http://numerical.recipes}.
In particular we assumed a tolerance of $10^{-12}$ during the integration and a tolerance of $10^{-6}$ for the root finding algorithm. This method improves both the precision and the efficiency of finding BH solutions, in particular we find that the matching between the ($M$, $S_{2-}$) and the ($h_1$, $f_1$) parameters is improved of a factor $10^4$ 
compared to previous results.\\
The location of the fitting point can be changed in order to improve the numerical stability 
of the system, although our results do not depent on its precise location. 
It is convenient to set $m=1$ so that the radial coordinate $r$, 
and the constants $M$, $S_{2-}$ in eq.(\ref{ag}) 
are  all  measured in units of $1/m$. 
By continuously changing the value of $r_H$ it is therefore possible to systematically explore the dependence
of the asymptotic parameters of the solution on the parameters $h_1$ and $f_1$ of the local expansion near the horizon. 
Moreover, once the convergence is achieved, we further shoot inward towards $r=0$ in 
order to characterize the behavior of the metric coefficients near the singularity. A single integration from $r\gg 1/m$ to $r=0$ typically requires less than $10s$, however the exploration of the full parameter space can take hours.

At last, the space of possible solutions is  described in Fig.(\ref{fig1}), where the inset on the right
shows the region around $r_H=0$. 
Black holes with $M>0$ only exist for $S_{2-}>-1.5$. Moreover, black holes with $S_{2-}>0$, represented 
with a blue line in the inset of Fig.(\ref{fig1}), are colder
than the Schwarzschild black hole with the same horizon radius, while black holes with $S_{2-}<0$ are instead hotter. If we compare the temperature of the non-Schwarzschild BH and the Schwarzschild ones with respect to the mass $M$, we find that the non-Schwarzschild ones are always colder. Black holes with large event horizon always have $M<0$. In the limit of zero temperature as $r_H\rightarrow 0$, the mass below $r_H<0.4$ assumes the constant value $M=M_0=0.62$. 
\begin{figure}[htpb]
\includegraphics[width=.48\textwidth]{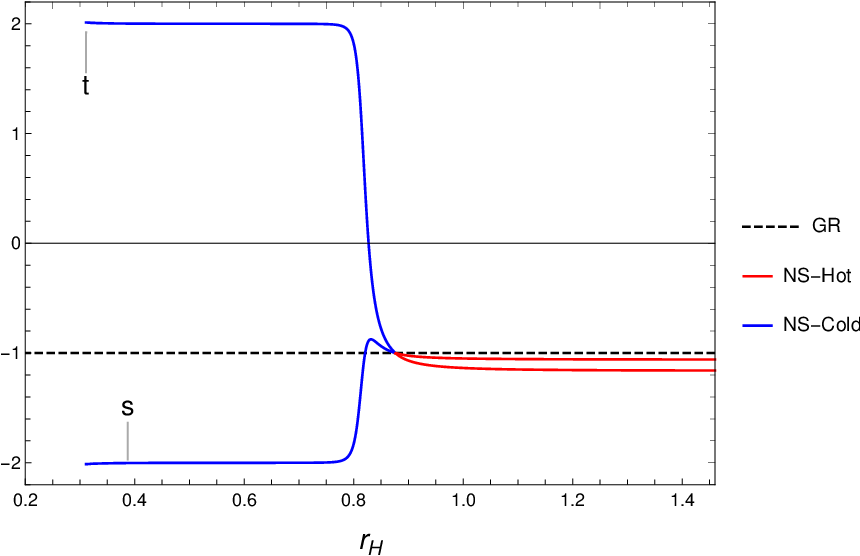}
\caption{Near-origin behaviour of the black hole solutions described in Fig.(\ref{fig1}).}
\label{fig3}
\end{figure}
\section{ The singularity at \texorpdfstring{$r=0$}{Lg}}
It is interesting to study the structure of the metric coefficient near $r=0$ as we move from the
hot branch to the cold branch in Fig.(\ref{fig1}).
Therefore for each value of $r_H$ we further integrate towards $r=0$  
in order to study the running exponents 
\be
\label{st}
t =r \partial_ r{\ln h(r}), \quad s = r \partial_r {\ln f(r})
\ee 
which can be determined by stopping the numerical integration at a limiting value
of the radius (we used $r=10^{-6}$ in order to preserve numerical stability).  
The results are depicted in Fig.(\ref{fig3}) where it can be noticed that as
$r_H$ runs from the hot branch to the cold branch, $(s,t)$ run from $(-1,-1)$ 
to $(-2,2)$ (note that in our notation $s$ has the opposite sign of the one in \cite{stelle2}). The limiting configuration reached in the $r_H\rightarrow 0$ limit
is the vanishing metric solution described in \cite{holdom}. 

To study this transition in detail let us define $x=-\ln r$ and rewrite eq.(\ref{uno}) and eq.(\ref{due}) as a function
of $s(x)$ and $t(x)$ in (\ref{st}). Exploiting the fact that $f$ is  large towards $r=0$ ($x=\infty$)
it is easy to obtain the following autonomous dynamical system
\begin{subequations}
\label{dyna1}
\be
 \frac{ds}{dx}= -\frac{-2 s^2 t+s^2-s t^2-8 s+t^3-3 t^2-8}{2 (t-2)}
\ee
\be
\frac{dt}{dx}=-\frac{1}{2} (-s t-4 s-t^2-2 t-4)\qquad\qquad\ \
\ee
\end{subequations}
Stationary solutions at $x=\infty$ determine the behavior of the metric near $r=0$. 
There are two fixed points (in addition to the trivial one $(0,0)$), $A=(-1,-1)$ which is an attractive improper node, and 
$B=(-2,2)$ which is an attractive node. Therefore as we move in the $(M,S_{2-})$ plane, the asymptotic behaviour
near $r=0$ is completely described either by $A$ or $B$ in complete agreement with the Frobenius analysis in \cite{stelle2}. 
We find that around $r_H=0.86$ a transition occurs, as shown in Fig.(\ref{fig3}), between a singular and a vanishing metric in the origin, in correspondence with the transition between the hot and cold branch. Note however that, due
to the improper node character of the point A, the approach towards $(-1,1)$ is much slower, as displayed in
Fig.(\ref{fig3}).

\section{Non-Schwarzschild black hole for \texorpdfstring{$m^2<0$}{Lg}}
Motivated by recent results on the Asymptotic Safety scenario, we now consider  the case $\alpha<0$ which implies
that the spin-2 mass is imaginary. In this case the large distance expansion reads
\begin{equation}
\label{at}
\begin{split}
&h(r)=1+C_t-\frac{2M}{r}+2 A_2\frac{\cos{\left(|m|r+\varphi\right)}}{r}\\
&f(r)=1-\frac{2M}{r}+A_2\frac{\cos{\left(|m|r+\varphi\right)}}{r}\, +\\
&+A_2|m|\sin{\left(|m|r+\varphi\right)}
\end{split}
\end{equation}
which depends on four unknown constants 
(two coefficients $M$ and $A_2$,  one phase $\varphi$, and the constant $C_t$ which 
we set to zero in the following).  The spacetime is no longer asymptotically flat and  we must require 
$A_2 |m|\ll 1$ for our linearized solution to be valid at large values of the radial coordinate $r$.

Although spatial oscillations of this type have been discussed before in the framework
of linearized gravity \citep{boos18}, in this work a complete solution
is presented for the first time in the framework of Einstein-Weyl theory.

\begin{figure}[htp]
\includegraphics[width=.48\textwidth]{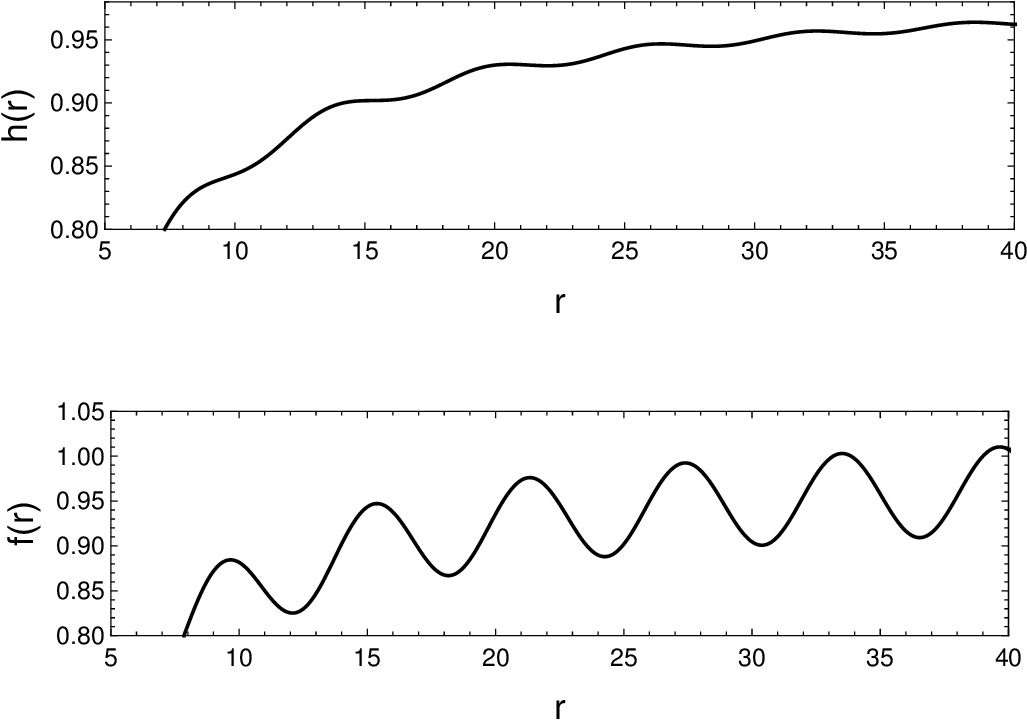}
\caption{A typical black hole solution for  $m^2<0$. Notice the non-asymptotically flat behavior for large $r$ in the bottom panel. For this solution we have $A_2=-0.051$ and $M=0.746$.
}
\label{fig4}
\end{figure}

The shooting method used for the $m^2 >0$ case can be applied also in this case.
However, if in the ghost case we imposed asymptotic flatness at infinity and found a one-parameter family of solutions, in this case we leave the phase $\varphi$ as free parameter and we find a two-parameter family. This is a reflection of the ill-defined limit of (\ref{at}) at large radii. A typical black hole solution is depicted in Fig.(\ref{fig3}) where one can notice the large $r$-behavior of the metric coefficient according to (\ref{at}). 

If we assume the condition $A_2\ll 1$ a resulting space of possible solutions is depicted in Fig.(\ref{fig4}). 
For the sake of clarity we show only some solutions with $0 < \varphi < \pi$; any solution with $\varphi' = \varphi + \pi$ has the same mass $M$, the same properties at the horizon but $A_2'=-A_2$. Smaller mass $M$ corresponds to smaller horizon radius, except for $\pi/2 < \varphi < \pi$, where the mass reaches a maximum and then decreases for increasing radius. Moreover, for any fixed phase, the solutions have a maximum horizon radius where the parameters $f_1$ and $h_1$ vanish and diverge, respectively.

\begin{figure}[htp]
\includegraphics[width=.48\textwidth]{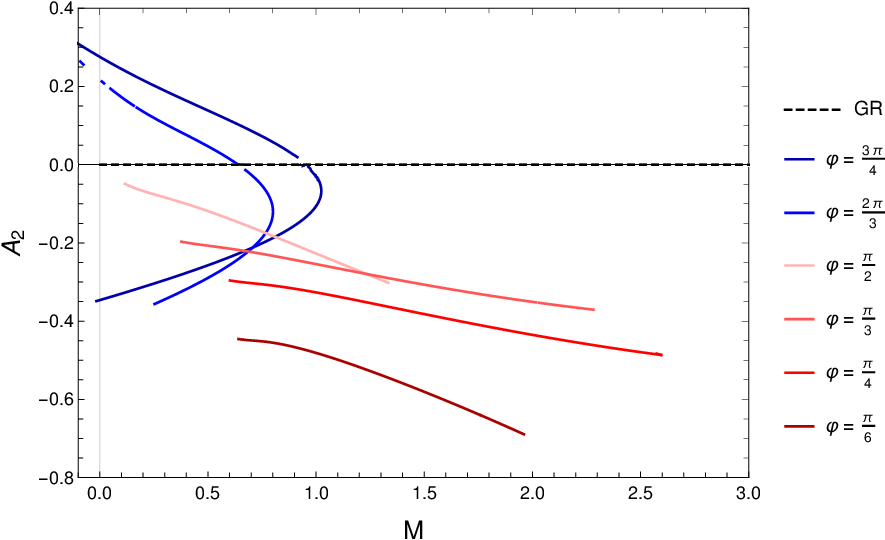}
\caption{Solutions space of Non-Schwarzschild black holes for $m^2<0$, matching the weak-field expansion in eq.(\ref{at}) at large $r$ with $0 < \varphi < \pi$, the solutions with $\pi < \varphi < 2\pi$ have the same $M$ but opposite $A_2$.}
\label{fig5}
\end{figure}
The solutions in blue are colder than the Schwarzschild solution with the same mass $M$, while the red ones are hotter; darker shades mean bigger differences between non-Schwarzschild and Schwarzschild temperatures. If we compare the temperatures in relation to the horizon radius, we find that BHs with $h_1<f_1$ are colder than the Schwarzschild ones, and the ones with $h_1>f_1$ are hotter. This phenomenon occurs for $A_2>0$ and $A_2<0$, respectively 
(for solutions with $0 < \varphi < \pi$).
\begin{figure}[htp]
\includegraphics[width=.48\textwidth]{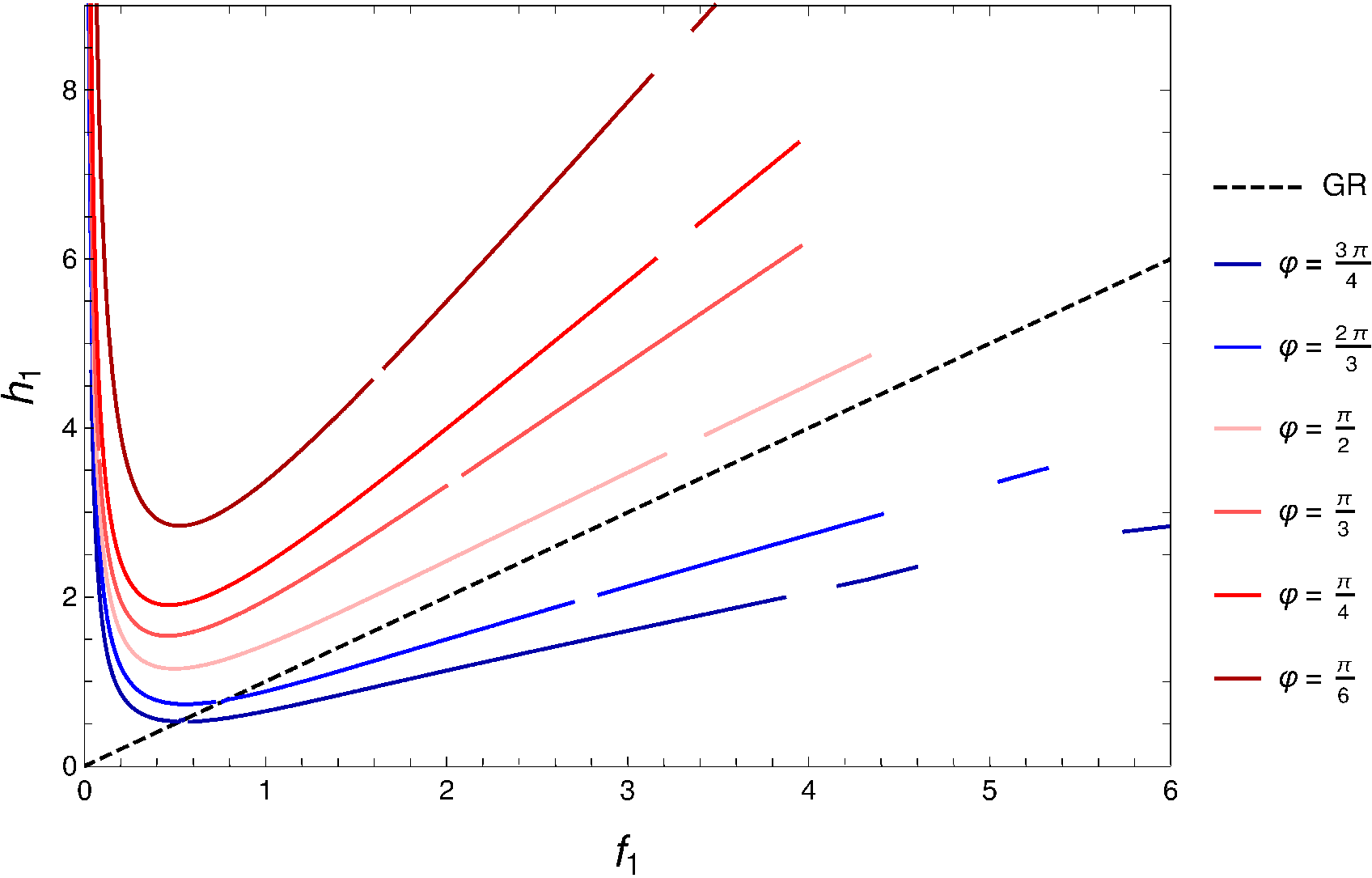}
\caption{Near-horizon parameters of Non-Schwarzschild black holes for $m^2<0$ with $0 < \varphi < \pi$; solutions with $\varphi'= \varphi+\pi$ have the same near-horizon parameters.}
\label{fig6}
\end{figure}
In Fig.(\ref{fig6}) we show the near-horizon parameter space, where increasing $f_1$ means decreasing $r_H$. It is interesting to note that the solutions populate almost the entire space, leading to a large variety in the thermodynamical properties of these black holes.
The near-origin behavior of this non-Schwarzschild black holes is still not clear, given that the increasingly oscillating nature of the solutions makes the additional inward integration numerically unstable.
\subsection{Conclusions}
In this work we clarified some issues in the study of black holes in quadratic gravity. We managed to link the asymptotic metric at large distances with the series expansion around the horizon. With this characterization it will be possible to study both the gravitational and thermodynamical properties of these new black holes and how these properties affect each other. 
Moreover, we made a first attempt to numerically study the interior of these solutions, finding that the cold, Yukawa-repulsive black holes have a vanishing metric in the origin while the hot, Yukawa-attractive ones have a singular metric. 

For the first time we analyzed also the case where the parameter $\alpha$ is negative
and then the massive spin-2 field is tachyonic instead of ghost-like. We found, together with Schwarzschild BHs, non-asymptotically flat black hole solution. 
It is interesting to notice that this behavior resembles the rippled structure of the vacuum found in $R+R^2$ gravity 
due to the kinetic condensation of the conformal factor\cite{bore13}.
A detailed study of the gravo-thermodynamical properties of both ghost-like and tachyonic non-Schwarzschild black holes will be presented 
in further works.

\subsection{Acknowledgements}
We would like to thank K. Stelle, A. Perkins, D. Litim and H. Gies 
for many stimulating discussions. One of us would like to thank the Osservatorio 
Astrofisico di Catania for its hospitality, the Universit\`a di Milano Bicocca for financial support, and A. Tomasiello for his constant support
and encouragement. We would also like to thank L. Santagati for careful reading of the manuscript.

%
\end{document}